\documentclass[journal]{IEEEtran}
\usepackage{blindtext}
\usepackage{graphicx}

\ifCLASSINFOpdf
\else
\fi
\begin{document}
%
\title{Mode Mixing Separation in Empirical Mode Decomposition of Signals with Spectral Proximity}
%
%

\author{Olav B. ~Fosso*,~\IEEEmembership{Senior Member,~IEEE,}
        Marta~Molinas*,~\IEEEmembership{Member,~IEEE,}
\thanks{O.B. Fosso is with the Department
of Electric Power Engineering, Norwegian University of Science and Technology (NTNU), 7491 Trondheim, Norway,
e-mail: olav.fosso@ntnu.no}
\thanks{M. Molinas is with Department of Engineering Cybernetics, Norwegian University of Science and Technology (NTNU), 7491 Trondheim Norway,e-mail: marta.molinas@ntnu.no}
\thanks{* These two authors contributed equally to the work.}
\space
\thanks{Manuscript received September 14, 2017}}

\maketitle

\begin{abstract}
The Empirical Mode Decomposition (EMD) is a signal analysis method that separates multi-component signals into single oscillatory modes called intrinsic mode functions (IMFs), each of which can generally be associated to a physical meaning of the process from which the signal is obtained. When the phenomena of mode mixing occur, as a result of the EMD sifting process, the IMFs can lose their physical meaning hindering the interpretation of the results of the analysis. In the paper,
   \emph{One or Two frequencies? The Empirical Mode
Decomposition Answers}, 
 Gabriel Rilling and Patrick Flandrin \cite{flandrin1} presented a rigorous mathematical analysis that explains how EMD behaves in the case of a composite two-tones signal and the amplitude and frequency ratios by which EMD will perform a good separation of tones. However, the authors did not propose a solution for separating the neighboring tones that will naturally remain mixed after an EMD. 
 In this paper, based on the findings by Rilling and Flandrin, a method that can separate neighbouring spectral components, that will naturally remain within a single IMF, is presented. This method is based on reversing the conditions by which mode mixing occurs and that were presented in the map by Rilling and Flandrin in the above mentioned paper. Numerical experiments with signals containing closely spaced spectral components shows the effective separation of modes that EMD can perform after this principle is applied. The results verify also the regimes presented in the theoretical analysis by  Rilling and Flandrin.  

\end{abstract}


%
\IEEEpeerreviewmaketitle

\section{Introduction}
Empirical Mode Decomposition (EMD) has since it was first proposed in \cite{huang1}, demonstrated its capabilities within many application areas. EMD's data driven approach does not require any specific assumptions behind the underlying model and is able to handle both non-linear and non-stationary signals. However, the algorithm has some limitations in identifying two closely spaced spectral tones and components appearing intermittently in the signal. In such cases the EMD will have trouble to separate the closely spaced tones and the intermittency. This paper addresses the problem of separation of closely spaced spectral tones and proposes a new method for separation. The structure of the paper is: the concept of mode mixing is discussed first, then some of the existing solutions are highlighted before the new method is presented. A case study based on a synthetic signal demonstrates the method's capabilities.
\subsection{What is Mode Mixing?}

Although Mode Mixing has not been strictly defined in the literature, it is known to happen as a result of the way in which the Empirical Mode Decomposition is designed to extract monocomponents from a multi-component signal.  
Only modes that clearly contribute with their own maxima and minima can be identified by the sifting process of the EMD. When a mode cannot clearly contribute with extremas, the EMD will not be able to separate the mode in a single IMF and the mode will remain mixed in another IMF.   
The paper in \cite{flandrin1}, provides restrictions on when it is possible to extract a tone from a composite two tones signals. The ratio of the amplitudes and of the frequencies of the individual components of the signal, will determine whether the EMD will be able to separate them in two different IMFs or whether they will be interpreted as one single IMF. 

In this paper, Mode Mixing is broadly categorized in two groups. Depending on the source at the origin, they can be originated by: 
\begin{itemize}
\item presence of closely spaced spectral components
\item presence if intermittence

\end{itemize}
In general, the reported literature offers solutions to the mode mixing caused by the presence of intermittence, while separation of closely spaced spectral tones has remained unsolved \cite{klingspor}.

This paper proposes a principle that when implemented improves the ability of the EMD to separate closely spaced spectral tones. In the following section, the main contributions to the solution of the mode mixing problem are discussed in light of their evolution in time. The new method is introduced within this perspective.

\subsection{Existing Mode Mixing Solutions}

Mode mixing, observed in the context of the Empirical Mode Decomposition, and caused by either intermittency of a signal component or by closely spaced spectral tones, is a well recognized limitation of the EMD method \cite{kaiser}\cite{flandrin1}\cite{wu1}. The principle behind the EMD is to always extract the highest frequency component present is the residue of the signal. In the presence of intermittence, the next lower frequency components may appear in the same IMF where the intermittent signal is identified, even though they are in different octaves and should appear as individual tones. In the case of closely spaced spectral tones, the signals will appear in the same IMF unless successfully separated. In 2005, in the paper: \emph{The Use of Masking Signal to Improve Empirical Mode Decomposition}, Ryan Deering and James F. Kaiser \cite{kaiser}  discuss the phenomena of mode mixing and presented for the first time the idea of using a masking signal to separate mixed tones caused by  the presence of intermittency in the signal. A formula for choosing the masking signal was suggested based on the observations of  empirical trials with several signals with mode mixing. The demonstration was essentially made for mode mixing caused by the presence of intermittency, since the frequency ratio between the signals was 0.57 which from the perspective of closely spaced spectral tones, was only moderately mixed. Rilling and Flandrin presented in 2008 a theoretical analysis that explains the behavior of the EMD in the presence of two closely spaced spectral components. Their work demonstrates which spectral components could be expected to be separated by the EMD based on their frequency and amplitude ratios. A \emph{Boundary Map} prepared by the authors, provides a visual indication of the efficiency of separation of two tones depending on their amplitude and frequency ratios. This work did not present a solution for the mode mixing problem caused by closely spaced spectral tones. It rather establishes restrictions on when it is possible, using EMD, to extract a tone from a composite two tones signals. 
In 2009, Wu and Huang presented the \emph{Ensemble Empirical Mode Decomposition (EEMD)} as a solution to cope with the mode mixing phenomena \cite{wu1}\cite{wu2}. Again, this approach was intended to solve the mode mixing caused by the presence of intermittent components in the signal. The principle behind EEMD is to average the modes obtained by EMD after several realizations of Gaussian white noise that are added to the original signal. After this work by Huang, several versions of the Ensemble EMD were proposed in the literature and the method is widely used \cite{flandrin2} but it is considered to be computationally expensive. 
Most recently, in April 2017, a patent application, \emph{System and Method of Conjugate Adaptive Conjugate Masking Empirical Mode Decomposition} filed by Norden Huang et. al \cite{huang2} discloses a method for directly processing an original signal into a plurality of mode functions. The invention claims to exclude the problem of mode mixing caused by an intermittent disturbance but does not apparently address the mode mixing caused by closely spaced spectral tones.

Although Deering and Kaiser introduced the idea of a masking signal, an open question remains on how to choose the frequency and the amplitude of the masking signal, to separate closely spaced spectral tones. 

The following section presents and discusses the principle that can enable EMD to separate these class of mode mixing. Based on the idea of the masking signal presented by Deering and Kaiser and the knowledge of the restrictions for closely spaced spectral tones presented by Rilling and Flandrin, a masking signal can be designed in order to reverse an existing mode mixing condition. 


\section{The Proposed Method}

The method is based on the original idea of Deering and Kaiser, of injecting a masking signal. The extension to this original idea is the way in which the masking signal is defined. In this work, the frequency and amplitude of the masking signal are chosen according to the boundary conditions presented in \cite{flandrin1}. The principle is explained in the following. 
Based on the boundary conditions presented in\cite{flandrin1}, the existing mode mixing conditions of a given signal can be reversed by adding a masking signal that can enforce a controlled artificial mode mixing with one of the signal's components, leaving the other free of mode mixing. 
If a masking signal of appropriate frequency and amplitude is added to the original signal, this masking signal will be able to attract one of the mixed signals but not the other. This  principle of attraction between closely spaced spectral tones, will create a new controlled artificial mode mixing, where one of the mixed signals is combined with the known masking signal. In this way, one of the originally mixed signals becomes free of mode mixing and comes as a pure mode (IMF) out of the EMD process. The controlled artificial mode mixing can be easily removed, since the masking signal is known.  

The step by step procedure for extracting IMF’s when modes are mixed, based on the proposed method in this paper is the following:
\begin{enumerate}
    \item Construct masking signal $x_m$ based on the new principle defined in this paper,
    \item Perform EMD on $x_+$ = $x$ + $x_m$ and obtain the IMF $y_+$. Similarly obtain $y_-$ from $x_- =x$ - $x_m$
    
    \item Define IMF as $y$ = ($y_+ + y_-)/2$
\end{enumerate}

Step 1 will require to obtain the frequency information from the original data. This information is obtained as explained in section II.B. 
\subsection{Mode Mixing of Closely Spaced Spectral Components}
In \cite{flandrin1}, it is pointed out that the amplitude and frequency ratios of the signal components are crucial for the understanding of the basic principles behind mode mixing. In the same work, a \emph{Boundary Condition Map} portrays the different regions where mode mixing is likely to occur as a function of the relative frequencies and amplitudes involved in the two signals. It can be observed from the map that in the region where the ratio between the frequencies involved are between 0 and 0.5, mode-mixing will not be observed for a range of amplitude ratios. Mode mixing is observed as the frequency ratio is higher than 0.67 and approaches 1.0, where mode mixing will always occur for all amplitude ratios. For the sake of clarity and to aid the discussions, the same \emph{Boundary Map} is reconstructed in this paper and shown in Figure 1.

\begin{figure}
    \centering
    {\includegraphics[width=3.5 in,clip, keepaspectratio]{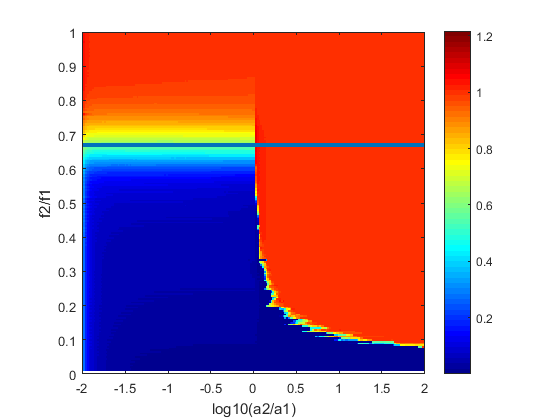}}
    \caption{Mode mixing boundary conditions map reproduced from \cite{flandrin1} used for defining masking signal}
    \label{fig:my-label}
\end{figure}

The map illustrates well how closely spaced spectral tones attract each other in a mode mixing. This same property is exploited in this paper for constructing effective masking signals to separate closely spaced spectral tones. 

\subsection{Principle for separation of closely spaced spectral components}

The principle applied in this paper is based on the combination of the technique presented by Kaiser and the \emph{Boundary Conditions Map} presented by Flandrin\cite{kaiser}\cite{flandrin1}.
The map of boundary conditions guides the choice of the masking signal's frequency and amplitude. 

To be able to extract a frequency by applying this principle, the ratio of that frequency to the frequency of the masking signal, must be located in the red area of the \emph{Boundary Map} (mode mixing area), while the ratio of the next frequency should be located in the blue area of the map. The amplitude ratios need to be adopted to ensure that condition as well. It is therefore necessary to operate with a frequency sufficiently close to the first mode and sufficiently distant from the next mode, to be successful. This criteria will be discussed in more detail in the case study.

Assume a signal with the two frequencies $f_1$ and $f_2$  ($f_1 > f_2$ ), where the  ratio between them will cause mode mixing due to close spectral tones. A masking signal of frequency $f_m$ larger than $f_1$ will attract $f_1$ if the ratio $f_1/f_m$ falls into the attraction region of the map (red color). If the ratio between $f_2$/$f_m$ falls in the region where there is no attraction (blue color) , adding a positive masking signal of frequency $f_m$ will separate the two signal $f_1$ and $f_2$ and the first IMF will have a controlled mode mixing of the signals $f_1$ and $f_m$. To separate $f_1$ and $f_m$  a negative masking signal may be added, and by averaging the two first IMFs the new IMF will be a signal of frequency $f_1$. However, depending on how close the two frequencies $f_1$ and $f_2$ are, some amplitude modulation will be observed between the signals $f_1$ and $f_2$.
To identify the  frequencies involved in the original signal, a Fast Fourier Transform (FFT) can be used as first screening tool. In this paper, a technique has been developed to identify the involved instantaneous frequencies and amplitudes, to assist in choosing the right masking signal.
Assume a signal $x$ defined by:
\begin{equation}
    x = A \sin(2\pi f_1 t) + B \sin(2 \pi f_2 t)
\end{equation}

After a standard EMD, these two signals will be mixed into one IMF. 
After a Hilbert-transform of the mode mixed IMF ($s = x + jy$) followed by an amplitude and an instantaneous frequency calculation, the required information is available for identifying the amplitudes and frequencies of the two signals involved. 
The instantaneous frequency used here is defined by: 
\begin{equation}
    f = \frac 1 {2\pi} \frac {\partial \phi} { \partial t}
\end{equation}
where:
\begin{equation}
    \tan \phi = \frac y x
\end{equation}
\begin{equation}
     \phi = \arctan \frac y x
\end{equation}

From the amplitudes of the Hilbert-transformed signal, the following expressions can be derived:
\begin{equation}
    K_{min} = \sqrt {A^2 + B^2 - 2AB} = (A - B)
\end{equation}
\begin{equation}
    K_{max} = \sqrt {A^2 + B^2 + 2AB} = (A + B)
\end{equation}

Similarly, the expressions for the extreme values of the instantaneous frequency plots are:
\begin{equation}
    F{min} = \frac {A \Delta f} {(A+B)} + f_2
\end{equation}
\begin{equation}
    F{max} = \frac {A \Delta f} {(A-B)} + f_2
\end{equation}

\noindent $K_{min}$ \indent{Minimum value of the amplitude plot}

\noindent $K_{max}$ \indent{Maximum value of the amplitude plot}

\noindent $F_{min}$ \indent{Minimum value of the instantaneous frequency plot}

\noindent $F_{max}$ \indent{Maximum value of the instantaneous frequency plot}

\noindent $\Delta f$  \indent{Difference between the two frequencies ($f_1 - f_2$)}\\

It is also demonstrated that $\Delta f$ is  equivalent to the number of peaks/second in the instantaneous frequency and the amplitude plots. The frequencies $f_1$ and $f_2$ can now be calculated and may be further validated with FFT calculation.

In the case of synthetic signals, these calculations are accurate and in principle the signal components could have been calculated directly by applying the above presented technique. However, for real signals the instantaneous amplitude and frequency functions are less smooth. Still they reveal information about the amplitudes and frequencies involved in the different periods of a mode mixed signal and from this, an optimal mask signal based on the map can be constructed.


\section{Case Study}
To demonstrate the method presented in this paper, a synthetic signal consisting of the following components is chosen:
\begin{equation}
    x = 0.7 \sin(2\pi 8 t) + 0.7 \sin(2 \pi 24 t) + 1.4 \sin(2 \pi 30 t) 
\end{equation}

The individual components and the final signal are shown in Figure 2.

\begin{figure}[ht]
    \centering
    {\includegraphics[width=3.5 in,clip, keepaspectratio]{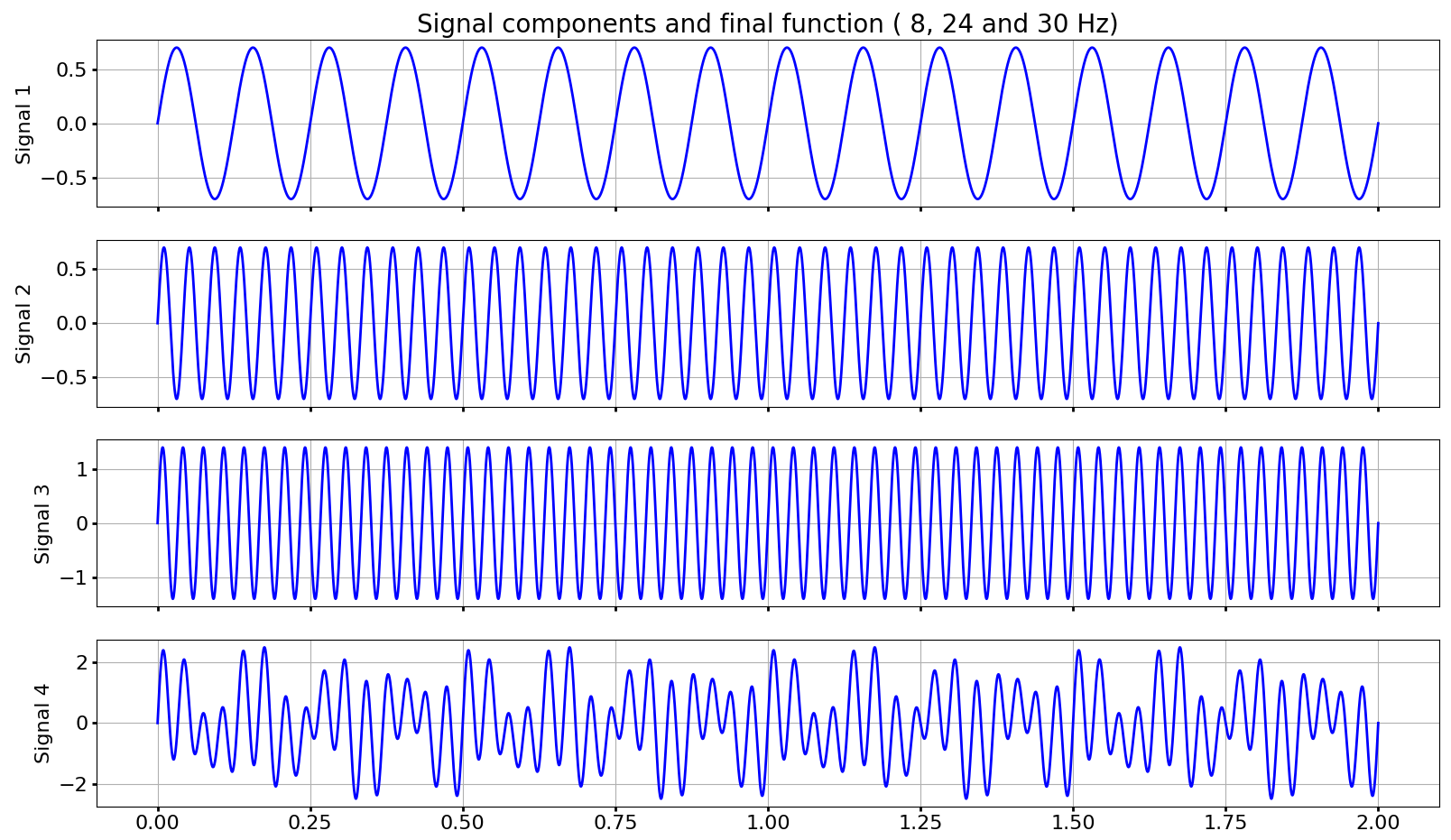}}
    \caption{Synthetic signal and its  components for case study}
    \label{fig:my-label}
\end{figure}

After separation with a standard EMD, the IMFs are shown in Figure 3. The first plot corresponds to the signal. The first IMF is the mix of the signals 24Hz and 30 Hz while the second IMF is the 8Hz signal well separated. The other IMFs reported are not relevant and only a result of the endpoint conditions, sifting process and applied tolerances.

\begin{figure}[ht]
    \centering
    {\includegraphics[width=3.5 in,clip, keepaspectratio]{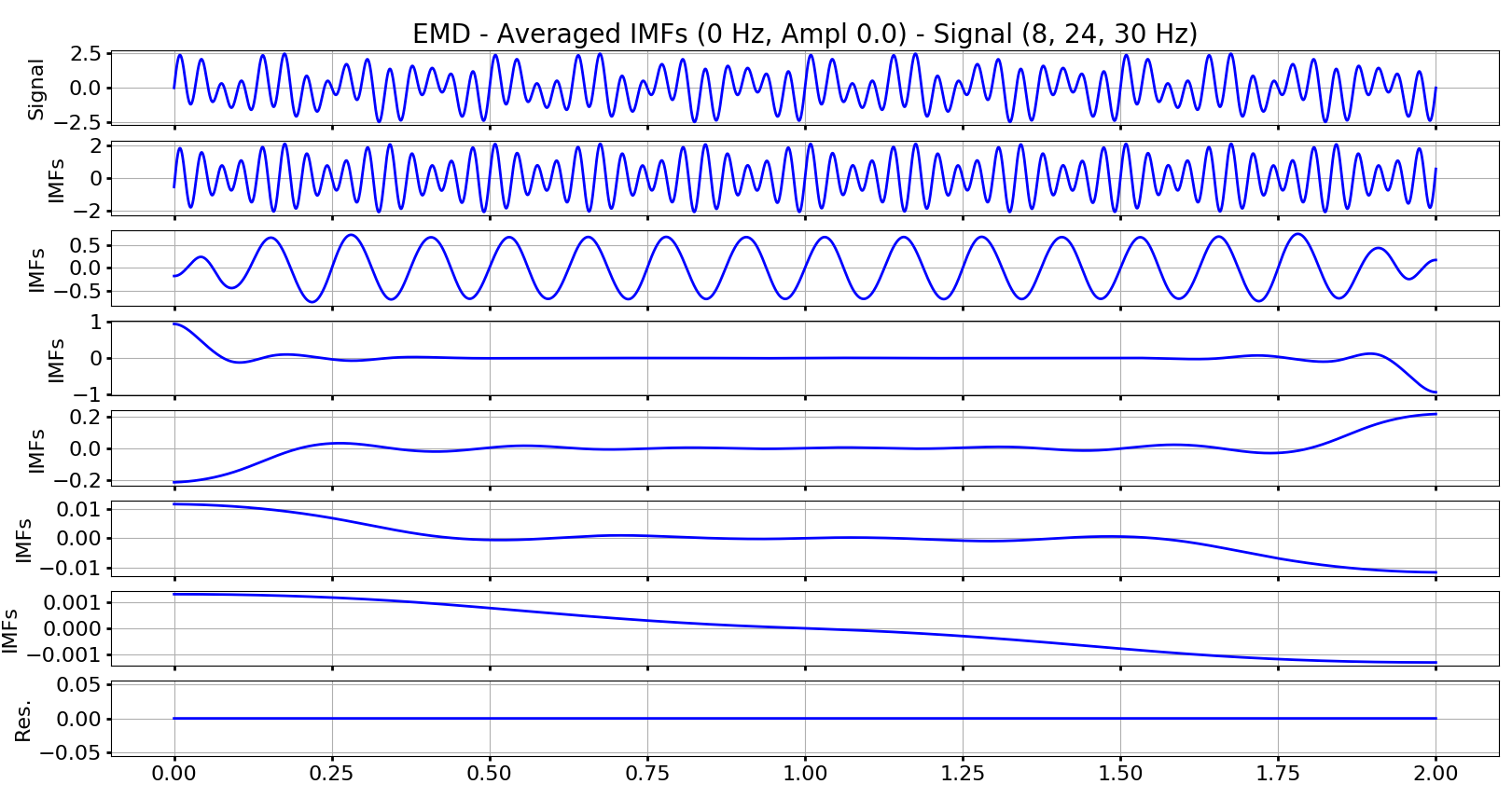}}
    \caption{Original signal, two IMFs and residuals after separation with a standard EMD}
    \label{fig:my-label}
\end{figure}
To identify the frequencies and amplitudes involved in the first IMF, a Hilbert transform is performed and the instantaneous amplitudes and frequencies are calculated. The instantaneous amplitudes are shown in Figure 4 and the instantaneous frequencies in Figure 5. Both figures portray the instantaneous values from the two IMFs.
\begin{figure}[ht]
    \centering
    {\includegraphics[width=3.5 in,clip, keepaspectratio]{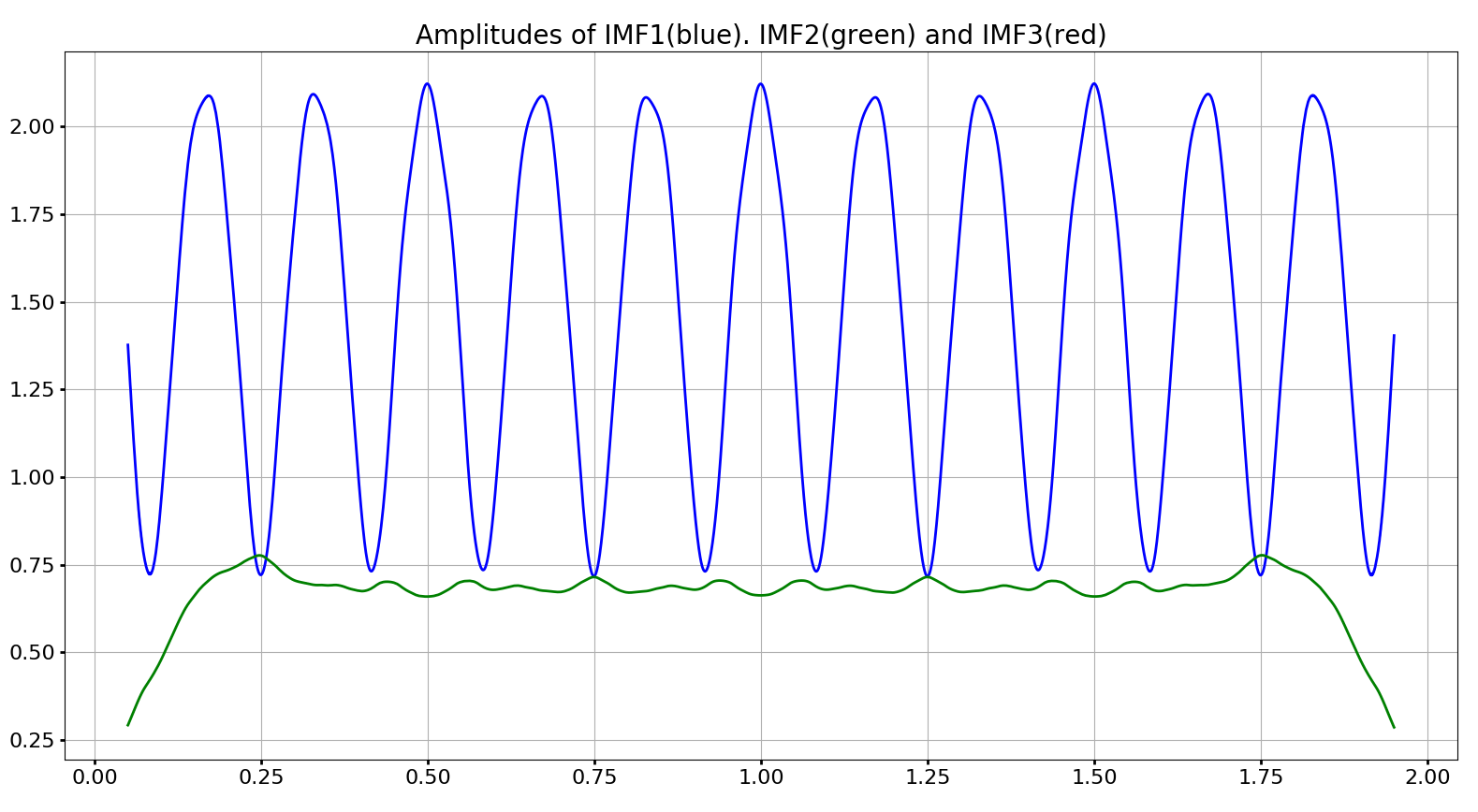}}
    \caption{Instantaneous amplitudes of IMF$_1$ and IMF$_2$}
    \label{fig:my-label}
\end{figure}
\begin{figure}[ht]
    \centering
    \includegraphics[width=3.5 in,clip, keepaspectratio]{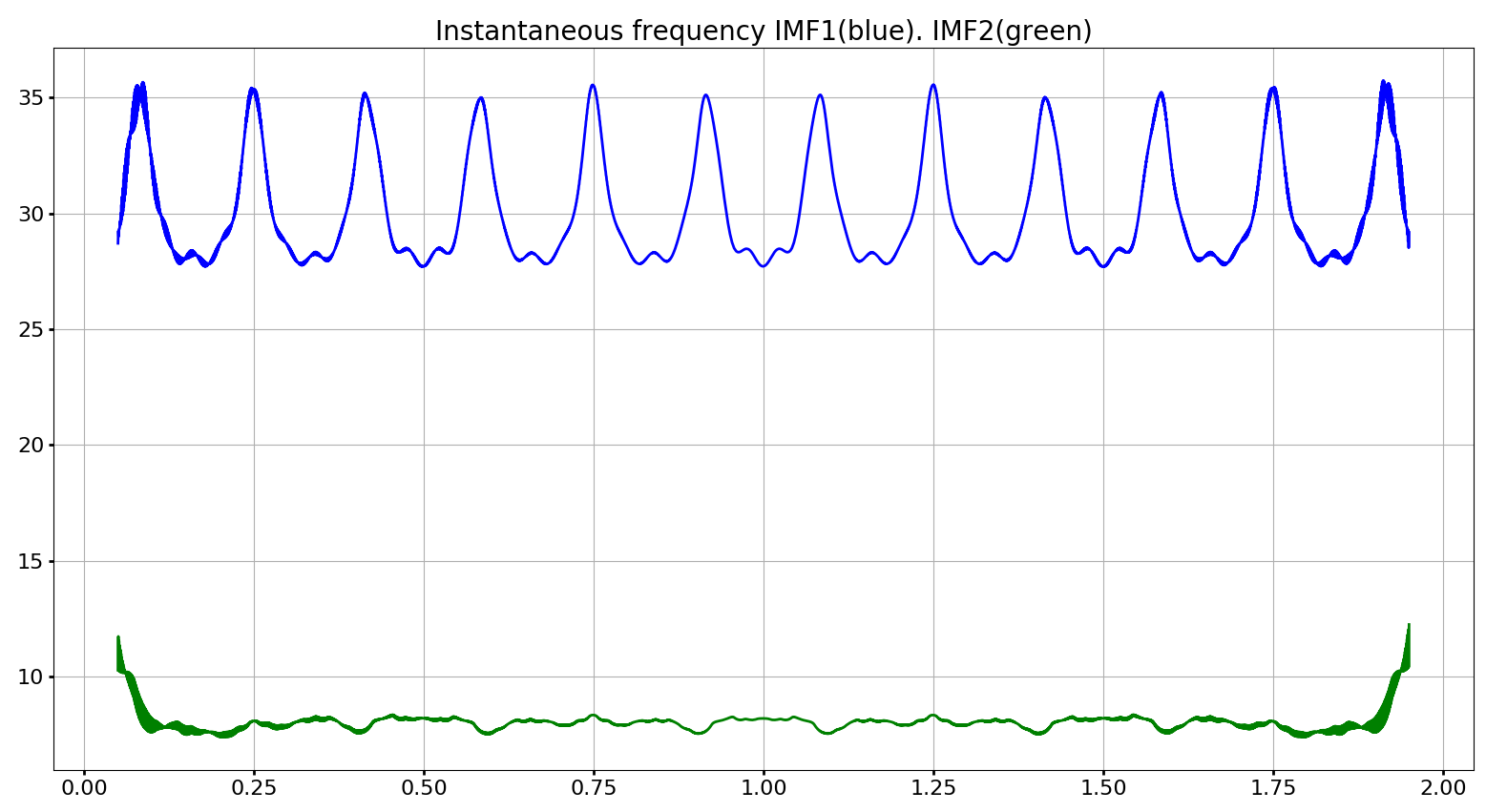}
    \caption{Instantaneous frequencies of the IMF$_1$ and IMF$_2$}
    \label{fig:my_label}
\end{figure}

To estimate the amplitudes and frequencies of the tones involved, we need to find the maximum and minimum of the amplitude function and of the instantaneous frequency function. Additionally we have to find the $\Delta f$ which is the number of \emph{peaks/second} in either the instantaneous amplitude or the instantaneous frequency plots. The maximum values extracted from the plots are indicated in Table 1. 

\begin{table}[h!]
\caption{Maximum values obtained from Inst. Amplitude and Inst. Frequency plots}
\begin{center}
\begin{tabular}{ |l|c|c|c| }
\hline
 \textbf{Frequency (Peaks/s)} & \textbf{$\Delta f$}  & 6 \\
 \hline
 \hline
 \textbf{Extreme values} & \textbf{Minimum} & \textbf{Maximum} \\
  \hline
  \textbf{Amplitudes} & 0.74 & 2.08 \\
  \hline
  \textbf{Instantaneous Frequencies} & 27.9 & 35.5 \\
  \hline
\end{tabular}
\end{center}
\end{table}

Using the obtained values in Table I into Equations 5-8, the estimated amplitudes and frequencies involved in the IMF with mode mixing are shown in Table II. These values are in good agreement with the original signal components.

\begin{table}[h!]
\caption{Estimated (Amplitudes / Frequencies) from Model}
\begin{center}
\begin{tabular}{ |l|c|c|c| } 
\hline
\textbf{Estimated parameters} & \textbf{Mode 1 (A and $f_1$)} & \textbf{Mode2 (B and $f_2$)}\\
  \hline
  \textbf{Amplitudes} & 1.41 & 0.67 \\
  \hline
  \textbf{Frequencies} & 30.06 & 24.06 \\
  \hline
\end{tabular}
\end{center}
\end{table}

Using the estimated values of A and B in Equation 8, the estimated frequency is: $f_2$ = 24.06. The process of extracting the IMFs depends on the number of iterations in the sifting process and may introduce some inaccuracies. This is verified by applying the technique directly on the signal, which gives exact values. 
For a synthetic signal, the separation could have been done directly. 
When applied in conjunction with EMD, the purpose is to extract amplitudes and frequencies for different parts of the signal to assist in constructing an optimal masking signal.

Based on the values obtained in Table II and following the recommendation of amplitude and frequency ratios observed in the \emph{Boundary Map} for mode mixing $f_1$/$f_m$ $>$ 0.67 (red/attraction area) and separation $f_2$/$f_m <<$ 0.67 (blue/separation area), an appropriate masking signal is chosen. 
Generally a good compromise is to keep $f_1$/$f_m$ $>$ 0.7 and  $f_2$/$f_m <$ 0.6 if possible. According to the map, a value of $a_m=2.5$ for the amplitude will be appropriate as $ log10 (a_1$/$a_m) = -0.25$ and $log10 (a_2$/$a_m) = -0.4$. With these values, IMF$_1$ after masking, will be located in the attraction area while IMF$_2$ in the separation areas (blue color).\newline

Now, using the chosen masking signal,  a new separation with EMD is conducted: 
\begin{equation}
    x_m=2.5cos(2\pi f_m t)
\end{equation}
Following the step by step procedure described in section II, the IMFs obtained with this procedures are shown in Figure 6.

\begin{figure}
    \centering
    {\includegraphics[width=3.5 in,clip, keepaspectratio]{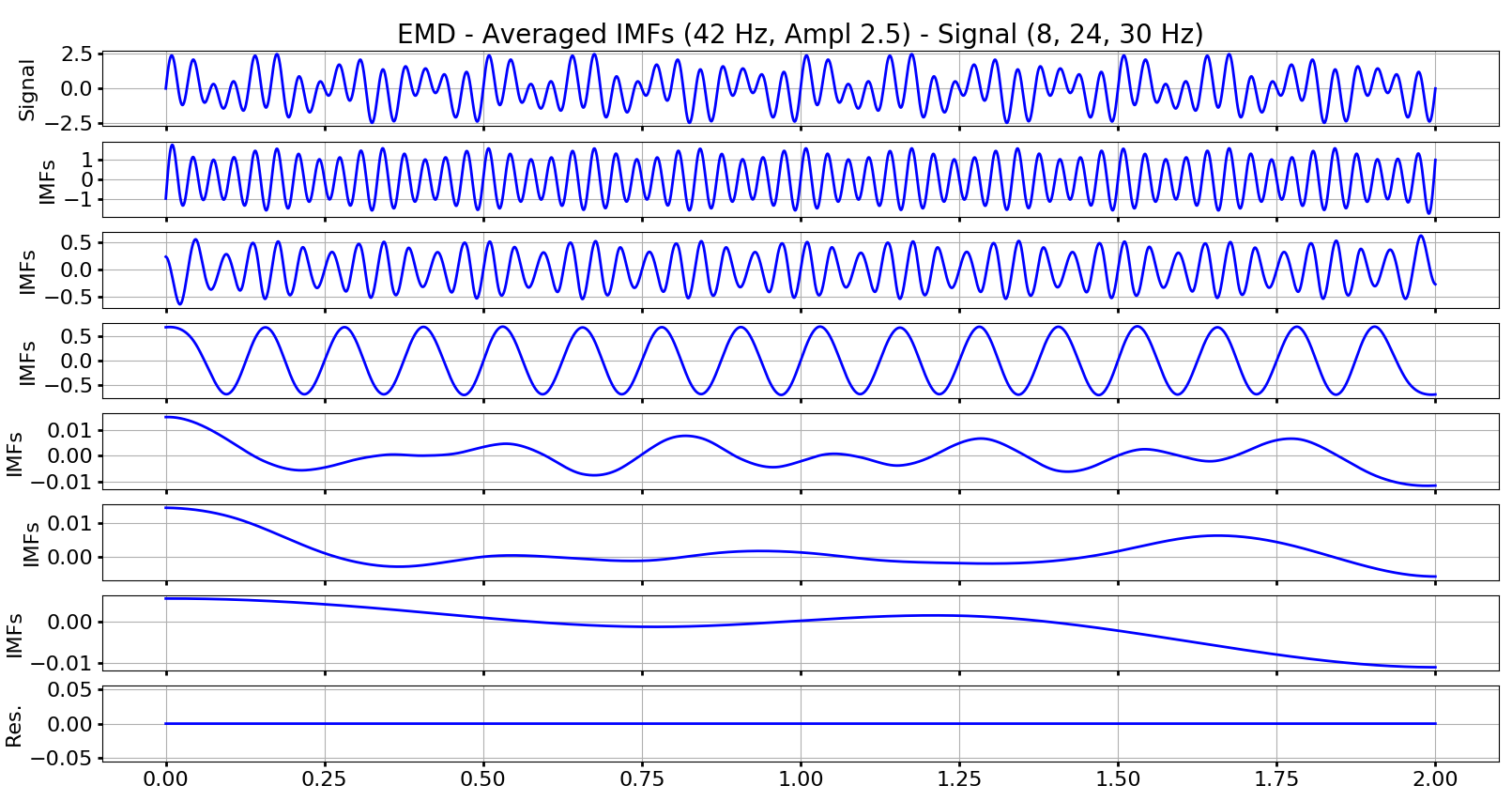}}
    \caption{IMFs after applying the masking signal defined by the proposed method}
    \label{fig:my-label}
\end{figure}

The three spectral components are now well separated. However, some amplitude modulation is observed on the 24 Hz and 30 Hz signals. When increasing the masking signal frequency, the amplitude modulation moves from the 30 Hz signal to the 24 Hz signal, while reducing the frequency will move the amplitude modulation to the 30Hz signal as the attraction is stronger according to the \emph{Boundary Map}.\newline

A Hilbert-transform is now performed on the two first IMFs followed by instantaneous amplitude and instantaneous frequency calculations. The results are shown in Figures 7 and 8. 

\begin{figure}
    \centering
    {\includegraphics[width=3.5 in,clip, keepaspectratio]{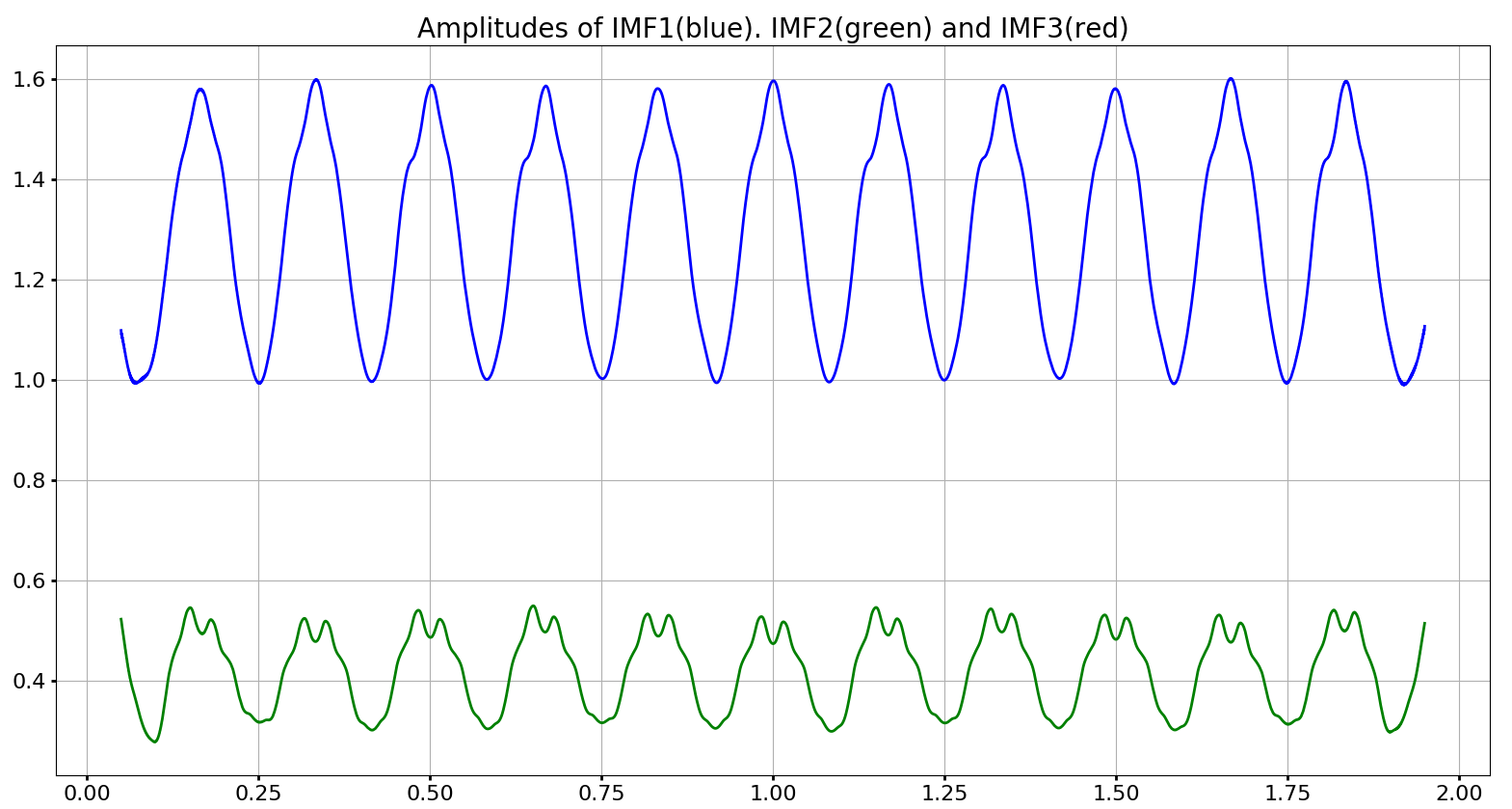}}
    \caption{Instantaneous amplitudes of IMF$_1$ and IMF$_2$ after proposed masking signal}
    \label{fig:my-label}
\end{figure}

\begin{figure}
    \centering
    {\includegraphics[width=3.5 in,clip, keepaspectratio]{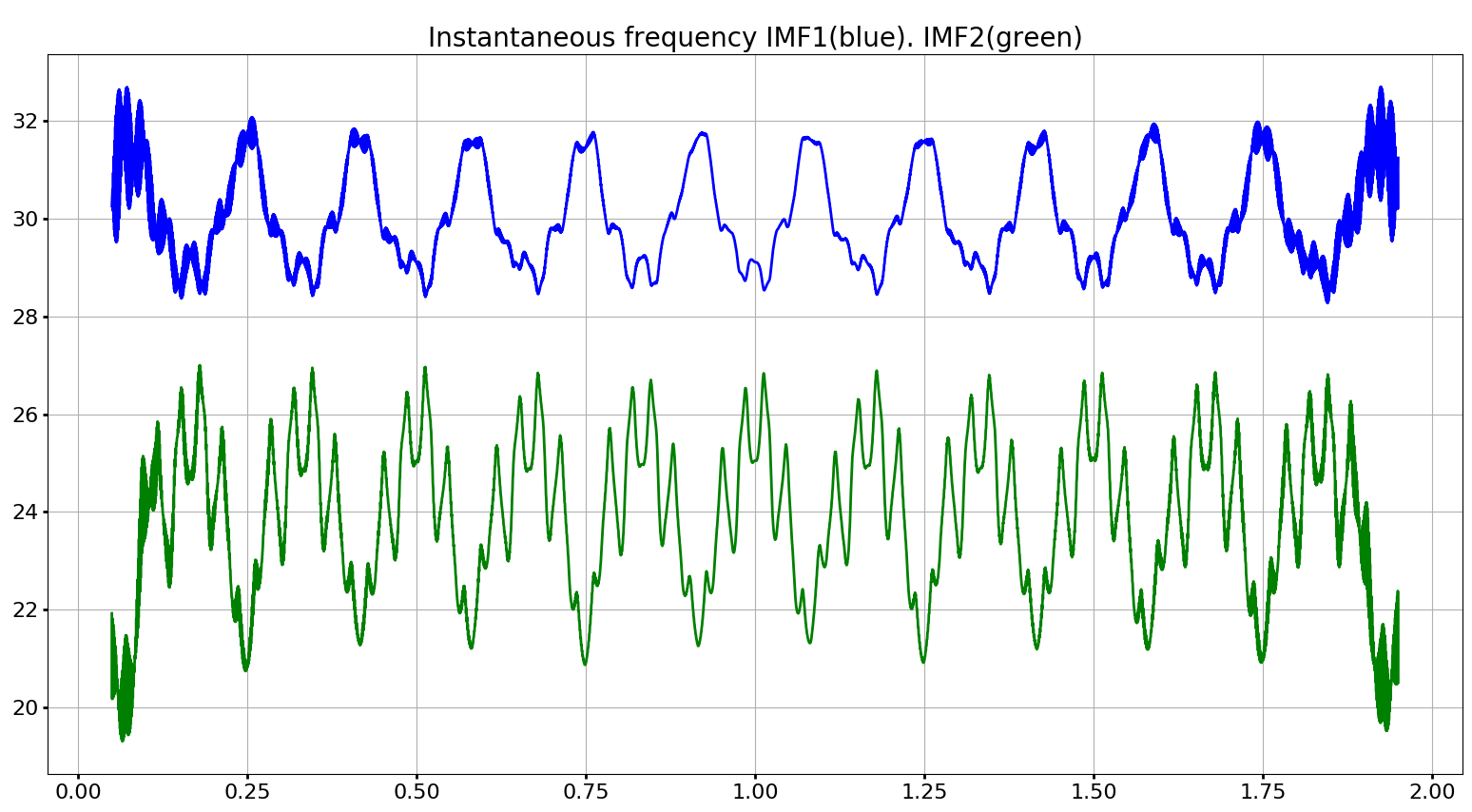}}
    \caption{ Instantaneous frequencies of  IMF$_1$ and IMF$_2$ after proposed masking signal}
    \label{fig:my-label}
\end{figure}

The instantaneous amplitude plots clearly shows the amplitude modulation and the presence of the other frequency component in both, IMF$_1$ and IMF$_2$. As the instantaneous frequency plots are less smooth, it is more difficult to identify the real extreme values needed in Equations 7 and 8. However, in this case we extract values from both IMF$_1$ and IMF$_2$. The extracted values are shown in Table III.

\begin{table}[h!]
\caption{Extreme values from Inst. Amplitude and Inst. Frequency plots}
\begin{center}
\begin{tabular}{ |l|c|c|c| }
\hline
 \textbf{Frequency (Peaks/s)} & \textbf{$\Delta f$}  & 6 \\
 \hline
 \hline
 \textbf{Extreme values} & \textbf{Minimum} & \textbf{Maximum} \\
  \hline
  \textbf{IMF$_1$: Amplitudes} & 1.0 & 1.6 \\
  \hline
  \textbf{IMF$_1$: Instantaneous Frequencies} & 28.8 & 31.79 \\
  \hline
  \hline
   \textbf{IMF$_2$: Amplitudes} & 0.32 & 0.55 \\
  \hline
  \textbf{IMF$_2$: Instantaneous Frequencies} & 21.4 & 26.9 \\
  \hline
\end{tabular}
\end{center}
\end{table}

Using the extracted values from Table III into Equations 5-8, the estimated amplitudes and frequencies involved in the IMF$_1$ and IMF$_2$ are shown in Table IV.

\begin{table}[h!]
\caption{Estimate (Amplitudes / Frequencies) from Model}
\begin{center}
\begin{tabular}{ |l|c|c|c| } 
\hline
\textbf{Estimated parameters} & \textbf{Mode 1 (A and $f_1$)} & \textbf{Mode2 (B and $f_2$)}\\
  \hline
  \textbf{IMF$_1$: Amplitudes} & 1.3 & 0.3 \\
  \hline
  \textbf{IMF$_1$: Frequencies} & 29.96 & 23.96 \\
  \hline
  \hline
   \textbf{IMF$_2$: Amplitudes} & 0.12 & 0.43 \\
  \hline
  \textbf{IMF$_2$: Frequencies} & 29.7 & 23.7 \\
  \hline
\end{tabular}
\end{center}
\end{table}

With reference to Equation 1, and the estimated values from Table IV, the IMF$_1$ and IMF$_2$ are estimated by:
\begin{equation}
    x_{1} = 1.3 \sin(2\pi 29.96 t) + 0.3 \sin(2 \pi 23.96 t)
\end{equation}
\begin{equation}
    x_{2} = 0.12 \sin(2\pi 29.96 t) + 0.43 \sin(2 \pi 23.96 t)
\end{equation}

The frequency components should be exactly the same (in Table IV) but since the extreme values of the instantaneous frequency were identified with low reliability, this can explain the difference. The plot of the estimates of the IMF$_1$ (red curve) together with the IMF$_1$ from the EMD (blue curve) are shown in Figure 9 for a period of 0.5 second. A very good agreement is also observed for the period of 2 seconds.

\begin{figure}[h!]
    \centering
    {\includegraphics[width=3.5 in,clip, keepaspectratio]{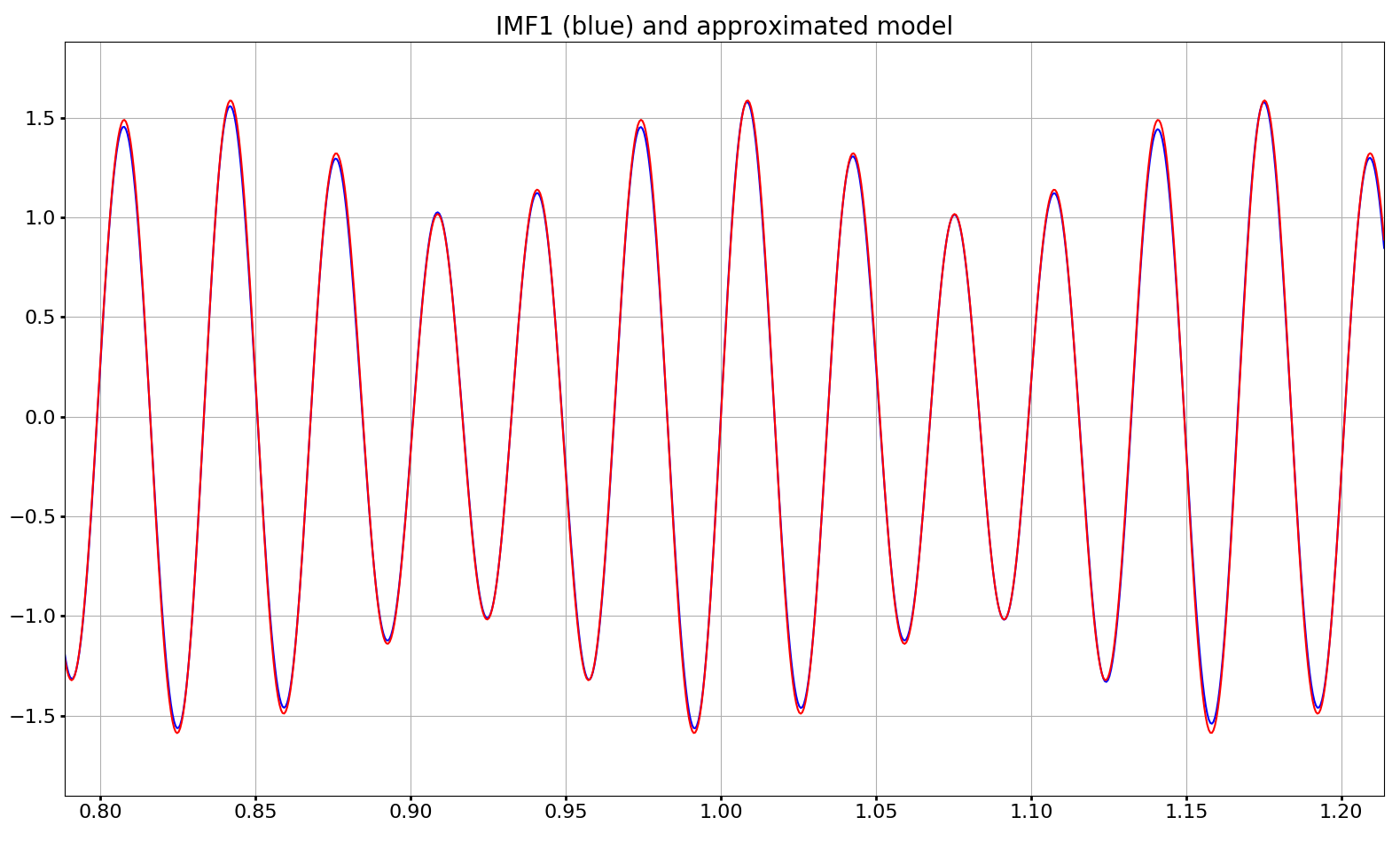}}
    \caption{ First IMF and estimated model - plot for 0.5 s}
    \label{fig:my-label}
\end{figure}

A similar plot is shown in Figure 10 for the IMF$_2$. 

\begin{figure}[h!]
    \centering
    {\includegraphics[width=3.5 in,clip, keepaspectratio]{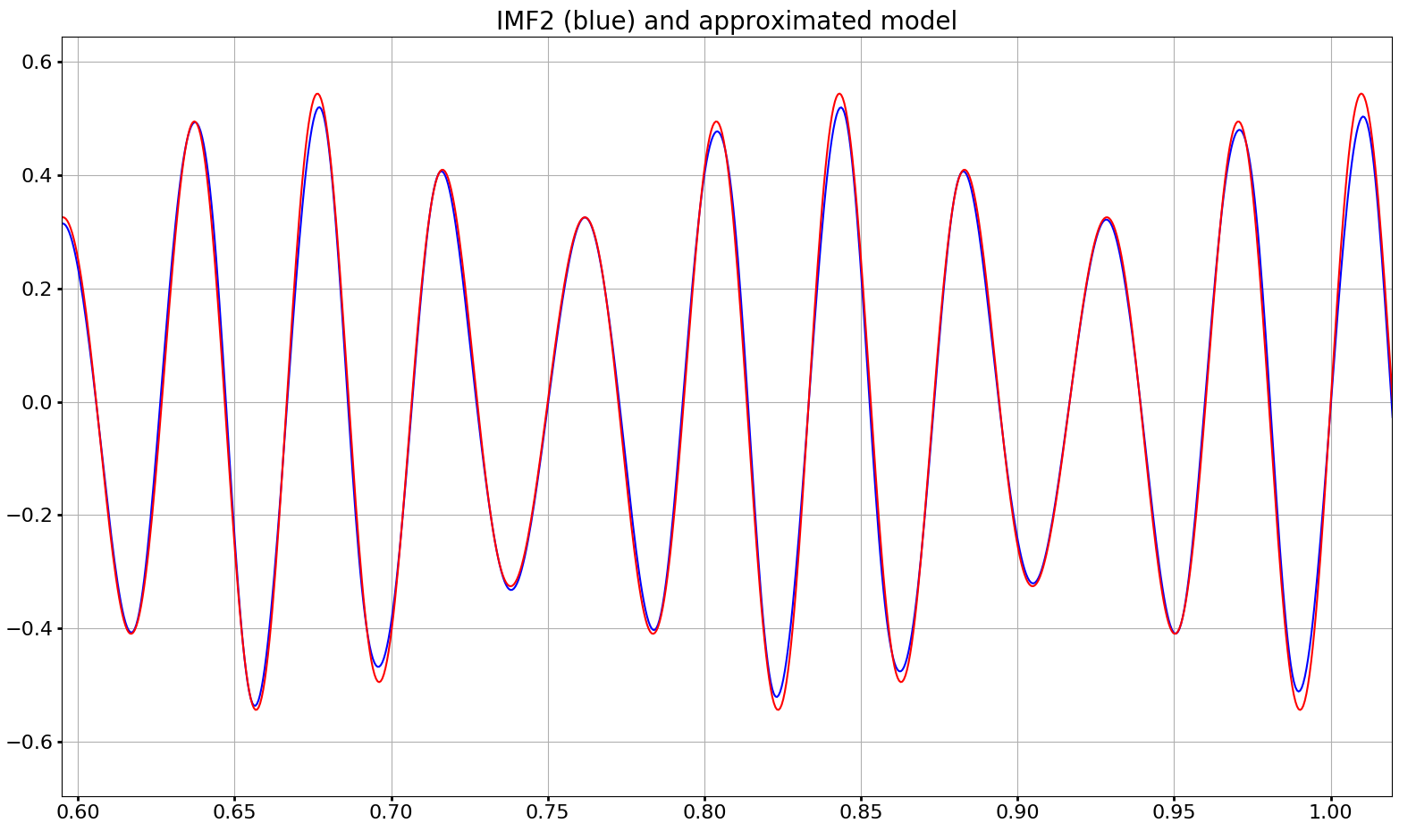}}
    \caption{ Second IMF and estimated model - plot for 0.5 seconds}
    \label{fig:my-label}
\end{figure}

For IMF$_2$, the agreement between the model and the IMF was also very good. In principle, the two corresponding components could have been added to get two pure sinusoidal functions of frequency 30 Hz and 24 Hz without any amplitude modulation. However, the phase shift of the sinusoidal functions are not accurately accounted for and therefore the conclusion could be misleading. 

\subsection{Summary of Findings}

This research aimed at enhancing the ability of the Empirical Mode Decomposition to separate closely spaced spectral tones. When two tones are in the same octave (the ratio between the $f_2/f_1 > 0.5$), the standard EMD will generate IMFs that are not mono-components. The main findings prove that by using the \emph{Boundary Map} presented in \cite{flandrin1}, it is possible to construct a masking signal which separates closely spectral tones. Even cases where the ratio $f_2/f_1$ exceeds 0.8, are separated with limited amplitude modulation. Further, it is found that using the instantaneous amplitudes and instantaneous frequencies obtained from the Hilbert-transform of the signal, it is possible to identify the frequencies and the amplitudes of the tones included in IMF with mode mixing. This information can be used to propose an optimal masking signal for separation of the closely spaced tones. In the case where the the signal cannot be separated without amplitude modulation, it is demonstrated how the same frequency and amplitude identification process can be used to find the amplitude and the frequencies of the amplitude modulated IMFs. This information may in specific cases be used to create mono-component IMFs.

\section{Discussion}
The work has essentially proposed a new method for separating closely spaced spectral tones using EMD. The principle was demonstrated and validated using synthetic signals. 
The model estimation process uses essentially local information obtained as a result of the EMD and Hilbert Transform process (number of peaks/s, extreme values of instantaneous amplitudes and instantaneous frequencies). The information is used to estimate the properties of the signal after a standard EMD has produced IMFs that may be mono-component or mixed mode IMFs. 

Further research in this topic is being dedicated to the definition of an optimal masking signal obtained from an optimization formulation aimed at minimizing the amplitude modulation.



%


\section*{Acknowledgment}

The authors would like to thank NTNU summer interns Francois Beline and Clara Renard for their contribution with the EMD study and the reconstruction of the \emph{Boundary Map} in \cite{flandrin1} used for defining the right value of the masking signal.

\ifCLASSOPTIONcaptionsoff
  \newpage
\fi




\begin{thebibliography}{1}

  
\bibitem{huang1} 
N. E. Huang, Z. Shen, S. R. Long, M. C. Wu, H. H. Shih, Q. Zheng, N.-
C. Yen, C. C. Tung, and H. H. Liu, \emph{The empirical mode decomposition
and the Hilbert spectrum for nonlinear and non-stationary time series
analysis}, Proc. of the Royal Society of London A: Math., Physical and
Engineering Sciences, vol. 454, no. 1971, pp. 903–995, 1998.

\bibitem{kaiser} 
R. Deering and J. F. Kaiser, \emph{The use of a masking signal to improve
empirical mode decomposition}, in Proc. IEEE Int. Conf. Acoust.,
Speech, Signal Process. (ICASSP ’05), 2005, vol. 4, pp. 18–23.

\bibitem{flandrin1} G. Rilling and P. Flandrin, \emph{One or two frequencies? The empirical
mode decomposition answers}, IEEE Trans. Signal Process., vol. 56,
no. 1, pp. 85–95, Jan. 2008.

\bibitem{wu1} 
Z. Wu, N. E. Huang, and X. Chen, \emph{The multi-dimensional ensemble
empirical mode decomposition method}, Adv. Adapt. Data Anal., vol.
1, no. 3, pp. 339–372, 2009.

\bibitem{wu2} 
Z.Wu and N. E. Huang, \emph{Ensemble empirical mode decomposition: A
noise-assisted data analysis method}, Adv. Adapt. Data Anal., vol. 1,
no. 1, pp. 1–41, 2009.

\bibitem{flandrin2} 
M.A. Colominas, G. Schlotthauer, M.E. Torres, P. Flandrin, 2012 : 
 \emph{Noise-assisted EMD Methods in Action}, 
 Adv. Adapt. Data Anal., Vol. 4, No. 4, pp. 1250025.1-1250025.11

\bibitem{huang2} 
Norden E. Huang, Zhao-Hua Wu and Jia-Rong Yeh, 
\emph{System and Method of Conjugate Adaptive Conjugate Masking Empirical Mode Decomposition}, U.S. Patent 2017/0116155 A1,
April 27, 2017 

\bibitem{klingspor}
Maans Klingspor: \emph{Hilbert transform: Mathematical theory and
applications to signal processing}, Thesis, University of Linkoping, November 2015, Electronic version: http://liu.diva-portal.org/smash/record.jsf?pid=diva2
\end{thebibliography}
%

%

\begin{IEEEbiography}[{\includegraphics[width=1in,height=1.25in,clip,keepaspectratio]{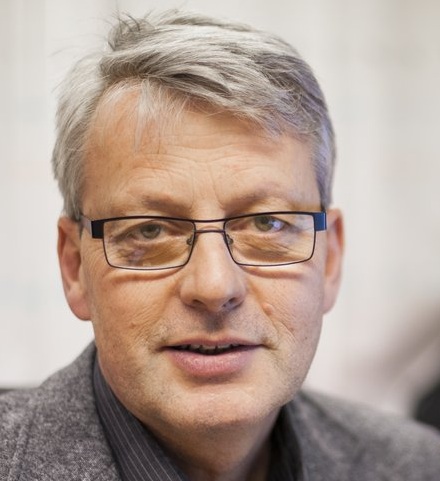}}]{Olav Bjarte Fosso}
is Professor at the Department of Electric Power Engineering of the Norwegian University of Science and Technology (NTNU). He has previously held positions as Scientific Advisor and Senior Research Scientist at SINTEF Energy Research, Head of the Department of Electric Power Engineering 2009 – 2013 and Director of NTNUs Strategic Thematic Area Energy from September 2014 - September 2016. He has been Chairman of CIGRE SC C5 Electricity Markets and Regulation and Member of CIGRE Technical Committee (2008 – 2014), Chairman of the board of NOWITECH (Norwegian Research Centre for Offshore Wind Technology) 2015-2017 and currently Board member of Energy21 (Norwegian National Strategy for Research, Development, Demonstration and Commercialization of New Energy Technology). He has been expert evaluator in Horizon2020 and in a number of science foundations, internationally. His research activities involve hydro scheduling, market integration of intermittent generation and signal analysis for study of power system’s dynamics and stability.
\end{IEEEbiography}

\vskip 0pt plus -1fil

\begin{IEEEbiography}
[{\includegraphics[width=1in,height=1.25in,clip,keepaspectratio]{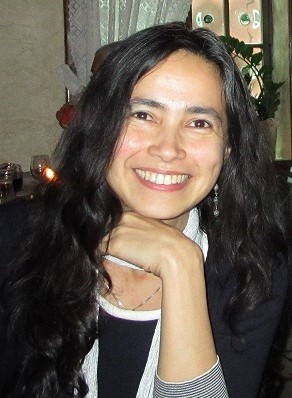}}]{Marta Molinas}
(M’97) received the Diploma degree in electromechanical engineering from the National University of Asunción, San Lorenzo, Paraguay, in 1992, the M.E. degree from the University of the Ryukyus, Nishihara, Japan, in 1997, and the D.Eng. degree from the Tokyo Institute of Technology, Tokyo, Japan, in 2000. She was a Guest Researcher with the University of Padova, Padua, Italy, in 1998. From 2004 to 2007, she was a Post-Doctoral Researcher with the Norwegian University of Science and Technology (NTNU), Trondheim, Norway. She was a JSPS Fellow at AIST in Tsukuba from 2008 to 2009. From 2008 to 2014, she was Professor with the NTNU Department of Electric Power Engineering. Since August 2014, she is a Professor with the NTNU Department of Engineering Cybernetics. Her current research interests include stability of power electronics systems, harmonics, instantaneous frequency, and non-stationary signals from the human and the machine.
Dr. Molinas is a Member of the IEEE Power Electronics Society, where she currently serves as an Associate Editor of the IEEE TRANSACTIONS ON POWER ELECTRONICS, the IEEE POWER ELECTRONICS LETTERS, and the IEEE JOURNAL OF EMERGING AND SELECTED TOPICS IN POWER ELECTRONICS. 
She has, in 2016, become Editor of the JOURNAL ON ADVANCES IN DATA SCIENCE AND ADAPTIVE ANALYSIS, World Scientific.
She is a member of the IEEE Industrial Electronics Society and Power Engineering Society.

\end{IEEEbiography}




\end{document}